\documentclass[11pt]{article}
\usepackage[margin=1.2in,a4paper]{geometry}
\usepackage{graphicx}
\usepackage{amsfonts}
\usepackage{amssymb}
\usepackage{amsmath}
\usepackage{latexsym}
\usepackage{subfloat}
\usepackage[small,sc]{caption2}

\title{\textbf{Metastable massive gravitons from \\ an infinite extra %
dimension\footnote{This essay received an honourable mention in the  2005
Gravity Research Foundation essay competition}}}

\author{Sanjeev S.~Seahra\footnote{Email: sanjeev.seahra@port.ac.uk} \\
Institute of Cosmology \& Gravitation \\ University of Portsmouth
\\ Portsmouth, PO1 2EG, UK }

\newcommand*{\di}{\partial}

\newcommand*{\Hank}[2]{\text{H}^{(#1)}_{#2}}

\begin{document}

\maketitle

\begin{abstract}

\setlength{\baselineskip}{15pt}

Motivated by stringy considerations, Randall \& Sundrum have
proposed a model where all the fields and particles of physics,
save gravity, are confined on a 4-dimension\-al brane embedded in
5-dimensional anti-deSitter space. Their scenario features a
stable bound state of bulk gravity waves and the brane that
reproduces standard general relativity. We demonstrate that in
addition to this zero-mode, there is also a discrete set of
metastable bound states that behave like massive 4-dimensional
gravitons which decay by tunneling into the bulk. These are
resonances of the bulk-brane system akin to black hole quasinormal
modes---as such, they give rise to the dominant corrections to
4-dimensional gravity. The phenomenology of braneworld
perturbations is greatly simplified when these resonant modes are
taken into account, which is illustrated by considering
gravitational radiation emitted from nearby sources and early
universe physics.

\end{abstract}

\newpage

\setlength{\baselineskip}{21pt}

String theory, one of the leading candidates for the `theory of
everything,' tells us that the universe has more than four
dimensions, but everyday experience seems to suggest otherwise.
Over the years, there have been several attempts to explain why
extra dimensions are hidden, including the conventional assumption
that they have a compact topology with radii on the order of the
Planck length.  But recently, certain developments in
non-perturbative string theory have provided alternatives to this
`Kaluza-Klein' compactifaction.  The key ingredient is so-called
$d$-branes, which are $(d+1)$-dimensional hypersurfaces on which
the standard model particles and fields can be consistently
confined.  This raises the possibility that the observable
universe is a 3-brane, i.e., we are living on a `braneworld'
\cite[review]{Maartens:2003tw}.

But in these scenarios, gravity is free to propagate in the full
higher-dimensional `bulk' manifold.  This is potentially
worrisome, since the force of gravity we are familiar with behaves
in an entirely 4-dimensional manner. For example, in a
higher-dimensional universe we would expect deviations from
Newton's inverse square law at large distances, and such
deviations have never been measured. We could again invoke
compactification to rescue the model, but there is a different
intriguing strategy: What if we were able to find scenarios where
the bulk graviton was, in some sense, \emph{dynamically} bound to
the brane? Could we then allow the extra dimensions to be
infinite?

\begin{figure}
\begin{center}
\includegraphics{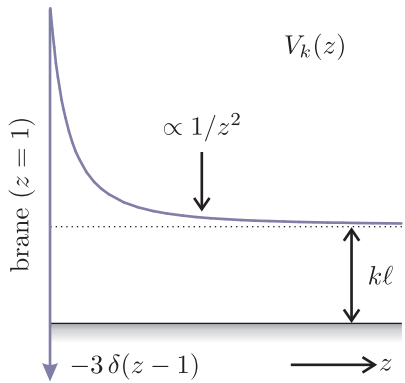}
\end{center}
\caption{The potential governing gravity waves in the
Randall-Sundrum braneworld}\label{fig:potential}
\end{figure}

In 1999, Randall \& Sundrum \cite{Randall:1999vf} proposed a
phenomenological realization of this idea in 5-dimensional
anti-deSitter space. Their model has the line element
\begin{equation}
    ds^2 = e^{-2|y|/\ell} \eta_{\alpha\beta} dx^\alpha dx^\beta + dy^2,
    \quad G^{(5)}_{ab} = (6/\ell^2) g^{(5)}_{ab},
\end{equation}
where $\ell$ is the anti-deSitter length scale. The brane is
identified with the intrinsically flat geometric defect at $y =
0$, and the Israel junction conditions imply that it is actually a
thin sheet of vacuum energy. Fluctuations of this model can be
written as $\eta_{\alpha\beta} \rightarrow \eta_{\alpha\beta} +
h_{\alpha\beta}$ with
\begin{equation}
    h_{\alpha\beta} =  e^{-|y|/2\ell}
    \psi_k(t,y) e^{-i\mathbf{k} \cdot \mathbf{x}} \epsilon_{\alpha\beta},
    \quad \epsilon_{\alpha\beta} = \text{constant},
\end{equation}
where $\psi_k$ satisfies a one-dimensional wave equation
\begin{equation}\label{wave equation}
    -\ell^2 \di_t^2 \psi_k = -\di_z^2 \psi_k + V_k(z) \psi_k, \quad z = e^{y/\ell} \ge 1.
\end{equation}
Here, we have restricted attention to one half of the bulk.  The
potential $V_k$ is shown in Figure \ref{fig:potential}. The delta
function enforces the boundary condition $\di_z (z^{3/2} \psi_k) =
0$ at $z = 1$, which ensures that the matter content of the brane
is unaltered by the perturbation. The attractive nature of the
delta function allows for a normalizable, stable bound state of
the brane and the 5-dimensional graviton
\begin{equation}
    \psi^{(0)}_k = e^{i\omega t} z^{-3/2}, \quad \omega = k.
\end{equation}
From the point of view of 4-dimensional brane observers, the
fluctuation behaves exactly like a massless spin-2 field
propagating on a flat background. Hence, this so-called
`zero-mode' reproduces 4-dimensional weak field gravity on the
brane, and shows how one can recover standard general relativity
with an infinite extra dimension.

However, $\psi^{(0)}_k$ is not the only solution to (\ref{wave
equation}).  There is a whole spectrum of modes parameterized by a
separation constant $m$:
\begin{equation}\label{massive mode solution}
    \psi^{(m)}_k = e^{i \omega t} \sqrt{z} \left[ \Hank{1}{1}(m\ell)
    \Hank{2}{2}(m\ell z) - \Hank{2}{1}(m\ell) \Hank{1}{2}(m\ell z) \right],
    \quad \omega^2 = k^2 + m^2,
\end{equation}
where $\Hank{n}{\nu}$ are Hankel functions.  The dispersion
relation on the right implies that each mode looks like a massive
spin-2 field to a brane observer.  So, in addition to the bound
state solution that reproduces general relativity, there is a
continuum of massive graviton `Kaluza-Klein' modes that predict
deviations from it. A complete description of brane gravity must
include both.

What is the essential behaviour of the massive modes on the brane,
and under what circumstances are they important relative to the
zero-mode? To be sure, a direct attack on this problem with
Green's functions can be mounted, but we pursue a more physical
approach by asking a different, more subtle question: Does the
potential shown in Figure \ref{fig:potential} treat each value of
$m$ democratically, or do certain masses tend to dominate the
gravity wave spectrum? That is, are any \emph{resonant} massive
modes within the Kaluza-Klein continuum? Resonant phenomena play a
prominent role in many branches of physics, and can often be used
to simply characterize the most important features of a given
system's dynamics.  There are at least two examples of this that
are useful to highlight: In black hole perturbation theory, the
master equations governing gravity wave propagation support
resonant solutions known as \emph{quasinormal modes}
\cite{Nollert}, which tend to dominate the late time behaviour of
scattered gravity waves, irrespective of the initial configuration
of the perturbation. A second example comes from accelerator
physics, where resonances in scattering cross sections allows one
to identify bound states of elementary particles \emph{without}
having a complete theoretical grasp on the interactions between
them. In both cases, we do not need to actually solve the
equations, subject to a given source, to predict the behaviour of
the system---the resonant effects are predominant. These model
problems suggest that the key to gaining physical intuition about
the massive modes in the braneworld is to find the resonant
solutions of (\ref{wave equation}).

Asymptotically far from the brane, the exact solution
(\ref{massive mode solution}) reduces to a superposition of
travelling waves, and the ratio of the coefficients of the
outgoing and incoming contributions defines the scattering matrix.
As in the two example problems cited above, the resonant modes of
our system are defined by the poles of this scattering
matrix.\footnote{For a comprehensive introduction to the theory of
one-dimensional scattering and the associated resonant phenomena,
see the textbooks by Taylor \cite{Taylor} or Landau and Lifshitz
\cite{LL}.} A simple calculation \cite{Seahra:2005wk} yields that
these correspond to a discrete set of \emph{complex} masses
\begin{equation}\label{resonant masses}
\{m_j\ell\} = \{\mu_j\} = \{0.419 + 0.577\,i,\, 3.832 +
0.355\,i,\, 7.016 + 0.350 \,i\, \cdots\}.
\end{equation}
If $k$ is real, the dispersion relation $\omega^2 = k^2 + m^2$
implies that these modes have $\text{Im}\,\omega > 0$, i.e., they
are exponentially damped in time.  In this sense, they are exactly
analogous to the quasinormal modes of perturbed black holes.
Because these solutions decay in time, it is sensible to call them
the metastable bound states of the brane and the bulk graviton. A
useful interpretation comes from recalling Gamov's classic 1928
treatment of the radioactive alpha decay \cite{Gasiorowicz}, where
the $\alpha$-particle is thought to be trapped in a potential well
surrounding a much heavier partner.  In that problem, the
metastable bound states represent solutions where the
$\alpha$-particle is mostly localized near the daughter nucleus,
but there is a finite probability of it tunnelling through the
potential barrier and out to infinity. In this spirit, we see that
the resonant masses (\ref{resonant masses}) represent a spectrum
of massive 4-dimensional gravitons mostly localized on the brane,
but subject to decay by tunnelling into the bulk.

This opens up a whole new way of looking at perturbative gravity
in the Randall-Sundrum scenario.  Before, one thought of a
zero-mass graviton accompanied by continuum of massive spin-2
fields.  Now, we realize that the essential behaviour is that of a
stable massless graviton augmented by a discrete family of
quasi-bound massive cousins.  To gain a better understanding of
the phenomenology of these `particles,' we consider a wavepacket
of gravitational radiation on the brane. We assume motion in the
$x$-direction, and a momentum space profile $\alpha(k)$. The
pulse's evolution on the brane will dominated by contributions
from the zero-mode and resonant masses:
\begin{equation}\label{pulses}
    \delta h_{\alpha\beta} \sim \epsilon_{\alpha\beta} \int dk \, \alpha(k) \left[
    \exp{ik(t-x)} + \sum_j c_j \exp{ik\left(\frac{t}{n_j}-x\right)} \right].
\end{equation}
Here, $j$ runs over the resonances, the $c_j$ expansion
coefficients are determined from the initial extra-dimensional
pulse profile, and $n_j$ is the complex reflective index
\cite{Jackson}
\begin{equation}
    n_j = n_j(k) = \frac{k}{\omega_j(k)} = \frac{k\ell}{\sqrt{(k\ell)^2+\mu_j^2}}.
\end{equation}
Hence, $h_{\alpha\beta}$ is a superposition of a discrete pulses
corresponding to the zero-mode and massive gravitons. Since $n_j$
has a nonzero real and imaginary parts, each of the massive pulses
behaves like it is travelling in an absorptive medium, slower than
the speed of light.  On the other hand, the zero mode acts like it
is propagating in a vacuum.

If $\alpha(k)$ is peaked about some $k = k_0$, we can sensibly
ask: How fast do the massive pulses travel? And how far do they
get before decaying?  We define the group velocity $v_j$ and
lifetime $\tau_j$ of each of the modes in the usual way:
\begin{equation}
    v_j = \text{Re} \, \omega_j'(k_0) = \text{Re}\,n_j(k_0),
    \quad \tau_j = \frac{1}
    {\text{Im} \, \omega_j(k_0)}
    = \frac{\text{Im}\,n_j(k_0)}{k_0}.
\end{equation}
Together, these give an attenuation length $d_j$, which is the
distance a given massive mode travels before its amplitude
decreases by a factor of $e$:
\begin{equation}
    \frac{d_j}{\ell} = \frac{v_j \tau_j}{\ell} =
    \frac{k_0\ell}{\text{Re}\,\mu_j\,\text{Im}\,\mu_j}.
\end{equation}
The denominator on the right is of order unity or larger, so we
see that modes with $k_0 \ell \gg 1$ can travel for large
distances, while modes with $k_0 \ell \ll 1$ have very short
streaming lengths on the brane.

Now, tabletop experiments of Newton's law limit $\ell \lesssim 0.1
\text{ mm}$, which immediately dashes any hopes of seeing any bulk
effects in the gravity waves emitted from nearby sources. The
reasoning is as follows: Astrophysical systems have sizes much
larger than 0.1 mm, which means that any gravitational radiation
will primarily be composed of partial waves with $k\ell \ll 1$.
Thus, the signal from the massive modes will only propagate for a
minuscule distance $d \ll \ell$ along the brane before decaying
away, making their direct detection impossible.\footnote{A
complimentary result can be derived by integrating over real
values of $\omega$ in (\ref{pulses}) and assuming that $k$ is
complex.  Then, one finds that the attenuation length becomes long
only when the source involves frequencies with $\omega\ell \gg 1$,
i.e., with $\omega \gg 10^{12} \text{ sec}^{-1}$.}

The situation is quite different when we consider cosmological
braneworld scenarios, where the motion of the brane in the extra
dimension accounts for the expansion of the universe.  In the
early universe epoch $H\ell \sim 1$ of such a model, all
sub-horizon modes will have $k\ell \gtrsim 1$.  Hence, the
attenuation length of the massive gravitons will be of order the
horizon size or larger, implying that they should have an
important effect of the gravity wave background. However, our
calculations to this point have been for a static brane, and it is
unclear how general brane motion affects the quasi-particle
excitations. But some (heuristic) progress can be made by
considering the small-scale fluctuations, which have
\begin{equation}\label{dispersion}
    k\ell \gg 1 \quad \Rightarrow \quad \omega_j(k)
    \approx k \left[ 1 + \frac{\mu_j^2}{2(k\ell)^2} \right]
    \quad \Rightarrow \quad |\omega_j(k)| \gg H.
\end{equation}
The last inequality means that the typical oscillation timescale
is much less than the speed of cosmological expansion, so it is
safe to say that these modes `see' the brane to be stationary, and
are still valid resonant solutions.  Hence, in the early universe
the high frequency component of the gravity wave background will
effectively be governed by a discrete spectrum of spin-2 fields
obeying the above dispersion relation. As the universe expands the
`fingerprint' of these fields expands with it, eventually
decoupling from the massive modes when its size grows larger than
$\ell$. Therefore, this `back of the envelope' calculation
predicts that the relic gravity wave background carries within it
primordial signatures of the extra dimension, encoded in
$\omega_j(k)$ and thereby the discrete set of complex numbers
$\{\mu_j\}$.

So, what has our knowledge of the resonances between bulk gravity
waves and the brane achieved for us?  We have seen that the late
time behaviour of metric fluctuations is dominated by a discrete
set of spin-2 fields with complex masses plus the zero-mode.
Kinematically, the resonant massive modes behave like gravitons
travelling in an absorptive medium---the dissipation is due to the
tunnelling of gravitational radiation into the extra dimension.
For a given AdS length scale $\ell$ in the bulk, we have
calculated the lifetime and streaming-length of these particles on
the brane, which are of order $\ell$ and $\ell/c$ respectively.
With $\ell \lesssim 0.1$ mm, we see that the massive part of the
spectrum cannot play a large role astrophysical processes in the
nearby universe, but may be significant in earlier epochs where
$H\ell \sim 1$.  This highlights the ephemeral nature of bulk
effects on the brane in the Randall-Sundrum scenario, and why such
a model is credible description of the \emph{real} world.  By and
large, we see that Einstein's theory of gravitation can---in
principle---live peacefully with infinite extra dimensions, and
the regions of conflict are neatly parameterized by discrete
spectrum of massive decaying gravitons.

I would like to thank Chris Clarkson, Roy Maartens, and David
Wands for helpful discussions and encouragement, and
\textsc{NSERC} for financial support.


\end{document}